\documentclass[aps,pra,twocolumn,groupedaddress,showpacs,floatfix]{revtex4-1}
\usepackage{amsmath}
\usepackage{amssymb}
\usepackage{physics}
\usepackage{mathtools}
\usepackage{mathrsfs}
\usepackage[pdftex,colorlinks=true,urlcolor=blue,linkcolor=blue,citecolor=blue]{hyperref}
\usepackage{graphicx}
\usepackage{float}
\usepackage[abs]{overpic}
\usepackage{pgfplots}
\pgfplotsset{compat=1.14}

\begin{document}

\title{The influence of sublattice bias on superfluid to Mott insulator transitions}

\author{Akshay Sawhney}
\email[E-mail: ]{as2755@cornell.edu}
\author{Erich J. Mueller}
\email[E-mail: ]{em256@cornell.edu}
\affiliation{Laboratory of Atomic and Solid State Physics, Cornell University, Ithaca, New York 14853, USA}

\date{\today}

\begin{abstract}
We model the superfluid to Mott insulator transition for a Bose gas on a lattice with two inequivalent sublattices.  Using the Gutzwiller ansatz, we produce phase diagrams and provide an understanding of the interplay between superfluidity on each sublattice.  We explore how the Mott lobes split, and describe the experimental signatures.
\end{abstract}


\maketitle

\section{Introduction}
The most iconic experiment in cold atom physics was the observation of the superfluid to Mott insulator transition \cite{greiner}.  In the superfluid phase atoms are delocalized throughout the entire lattice, while in the Mott insulating phase interactions suppress atomic motion.  This transition is a prototype for understanding strong correlation effects in quantum degenerate matter.  In subsequent years more sophisticated experiments were developed for exploring this transition \cite{demarco2005,campbell2006,folling2006,gerbier2005a,gerbier2005b,orzel2001,hadzibabic2004,schori2004,xu2006,sengupta2005,stoferle2004,campbell2006,gerbier2006,greiner2010}.  Of particular interest is the role lattice geometry, which is being explored in experiments which create exotic non-Bravais lattices \cite{stamperkurn}.  Here we conduct a mean-field study of interacting bosons in such lattices.

At its core the superfluid-Mott transition is a competition between kinetic energy, which favors macroscopically occupying delocalized states, and interactions, which favor product states where each site has a definite occupation.  As one changes the lattice geometry, the nature of the delocalized states can change, influencing this energy balance.  One extreme example of this behavior are studies of the superfluid-Mott transition in Hubbard-Hoffstadter models, which boast relatively flat bands with topological character \cite{Natu2016, Goldbaum2008, Umucallar2007, Umucallar2010}.  Here we explore a different phenomenon namely how the structure within a unit cell influences the transition.

\begin{figure}[tbhp]
    \centering
    \includegraphics[width=\columnwidth]{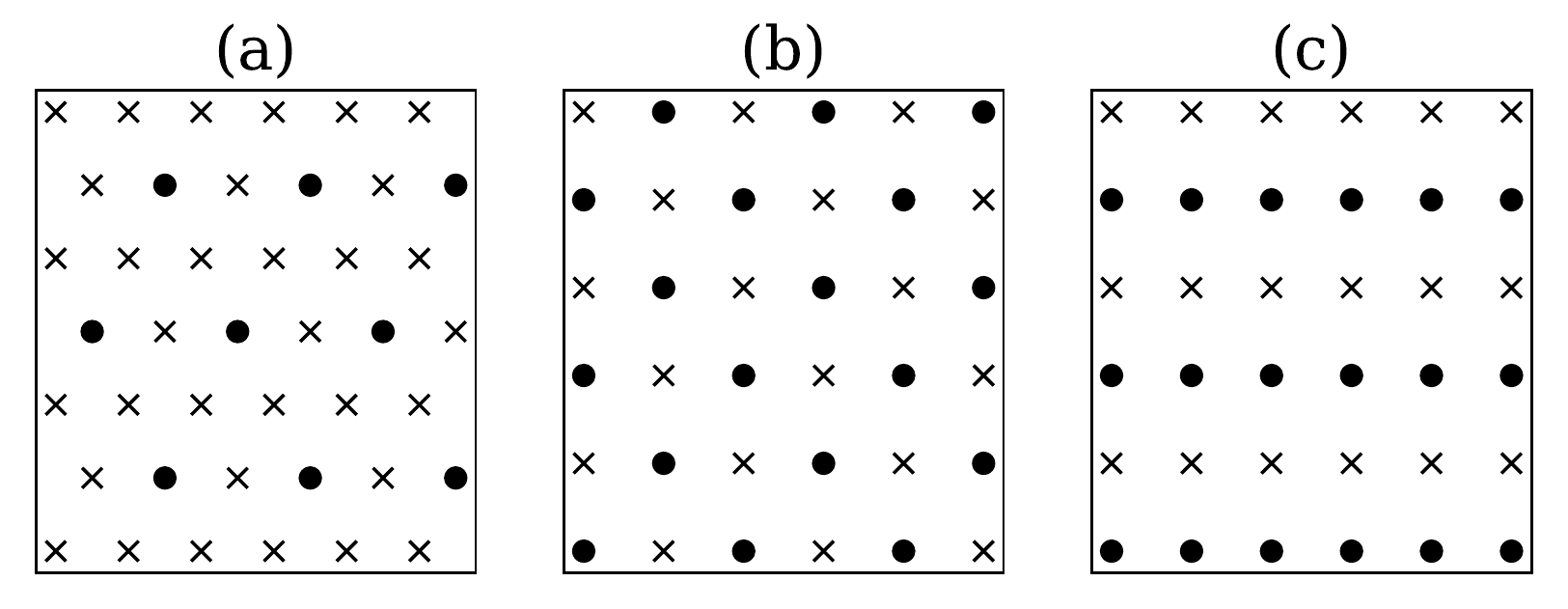}
  \caption{Example lattices: (a) Kagome (b) Checkerboard (c) Stripe.  In all cases the local potential is different on sites marked by x's and dots.}
  \label{fig:lattices}
\end{figure}

Motivated by experiments at Berkeley \cite{stamperkurn2017}, we consider the case where the lattice depth is different for two sublattices.  Figure~\ref{fig:lattices} shows some examples:  atoms sit at the sites marked by x's and dots, but have different energies on each of these sublattices.  The Berkeley experiments were restricted to the extreme cases where the energy offset was very large, and where it vanished.  Our work addresses the general problem: How does the phase diagram evolve as one changes the offset?
 
Technically, one creates such optical lattices by interfering laser beams.  The lattice geometry is set by the intensity and wave-vectors of the lasers.  Experiments have demonstrated kagome, checkerboard, striped, triangular and honeycomb lattices \cite{Becker2010,SoltanPanahi2011,Windpassinger2013,Tarruell2012,stamperkurn}.

\section{Model}
We consider bosons hopping on a lattice with two different types of sites, $A$ and $B$.  We will mostly be thinking about two-dimensional examples, but the formalism also applies to higher dimensions.  Extensions to more sublattices are straightforward, and most of the structure is clear from the two-sublattice case. 

Generically there will be terms in the Hamiltonian which involve $A$-sites, terms involving $B$-sites, and those that connect the two sublattices.  We will take the interactions to be strictly local, as is appropriate for most cold-gas experiments.  Thus we write $ \mathcal{H} = \mathcal{H}_A + \mathcal{H}_B + \mathcal{H}_{AB}$ with
\begin{gather}
\nonumber
\mathcal{H}_A =
    \sum_{i\in A}(
    \frac{U_A}{2}a_i^\dagger a_i^\dagger a_i a_i-\mu_A
    a_i^\dagger a_i)-t_A \sum_{\mathclap{\expval{i, j} \atop i,j\in A}}(a_i^\dagger a_j + a_j^\dagger a_i)
    \\ \nonumber
\mathcal{H}_B =
\sum_{i\in B} (
\frac{U_B}{2}
b_i^\dagger b_i^\dagger b_i b_i-\mu_Bb_i^\dagger b_i)
    -t_B \sum_{\mathclap{\expval{i, j} \atop i,j\in B}}(b_i^\dagger b_j + b_j^\dagger b_i)
\\
\mathcal{H}_{AB} =  -t_{AB}\sum_{\mathclap{\expval{i, j} \atop i\in A, j\in B}}(a_i^\dagger b_j + b_j^\dagger a_i)
\label{ham}
\end{gather}


Where $a_i^\dagger$ and $a_i$ are the creation and annhilation operators for the $A$ sites, $b_i^\dagger$ and $b_i$ are those for $B$ sites, and $\expval{i,j}$ denotes neighbors.  The coefficients $t_A,t_B,t_{AB}$ are hopping matrix elements, $U_A,U_B$ are on-site interactions, and $\mu_A,\mu_B$ represent the chemical potentials of the two sublattices.  In particular, for the experiments we are considering $\mu_B=\mu_A+V$, where $V$ is the energy offset.

The relevance of the various terms depends on the lattice geometry.  For example, the kagome lattice in Fig.~\ref{fig:lattices}(a) has no B-B neighbors, while the Checkerboard lattice in Fig.~\ref{fig:lattices}(b) has neither A-A neighbors nor B-B neighbors. 




As  explained by Fisher et al. \cite{fisher}, the
principle features of Bose-Hubbard models such as Eq.~(\ref{ham})
are captured by
the variational ansatz,
\begin{align}
    \ket{\psi} = \left(\bigotimes_{i\in A} \sum_n f_n^A\ket{n}_i\right)\left(\bigotimes_{i\in B} \sum_n f_n^B\ket{n}_i\right)
\end{align}
where $\ket{n}_i$ is the Fock state with $n$ particles at site $i$.  We characterize the ground state by minimizing $\langle \mathcal{H}\rangle=\bra{\psi}\mathcal{H}\ket{\psi}$, treating the $f_n^\sigma$ as variational parameters, with the normalization constraint $\sum_n |f_n^\sigma|^2=1$.

The order parameters for superfluidity in each sublattice are
\begin{align}\label{alphas}
    \alpha &\equiv \expval{a_i} = \sum_n \sqrt{n}\;f_{n-1}^A\left(f_n^{A}\right)^* \\
    \beta  &\equiv \expval{b_i} = \sum_n \sqrt{n}\; f_{n-1}^B \left(f_n^{B}\right)^* \nonumber
\end{align}
The hopping in $\mathcal{H}_{AB}$ couples these order parameters, so that if one of these is non-zero, then so is the other. More formally, $\langle \mathcal{H}\rangle$ contains a term proportional to $\alpha \beta$, implying that the only stationary point of the energy with $\alpha=0$ also has $\beta=0$.

\begin{figure*}[tbhp]
  \centering
  \includegraphics[width=\linewidth]{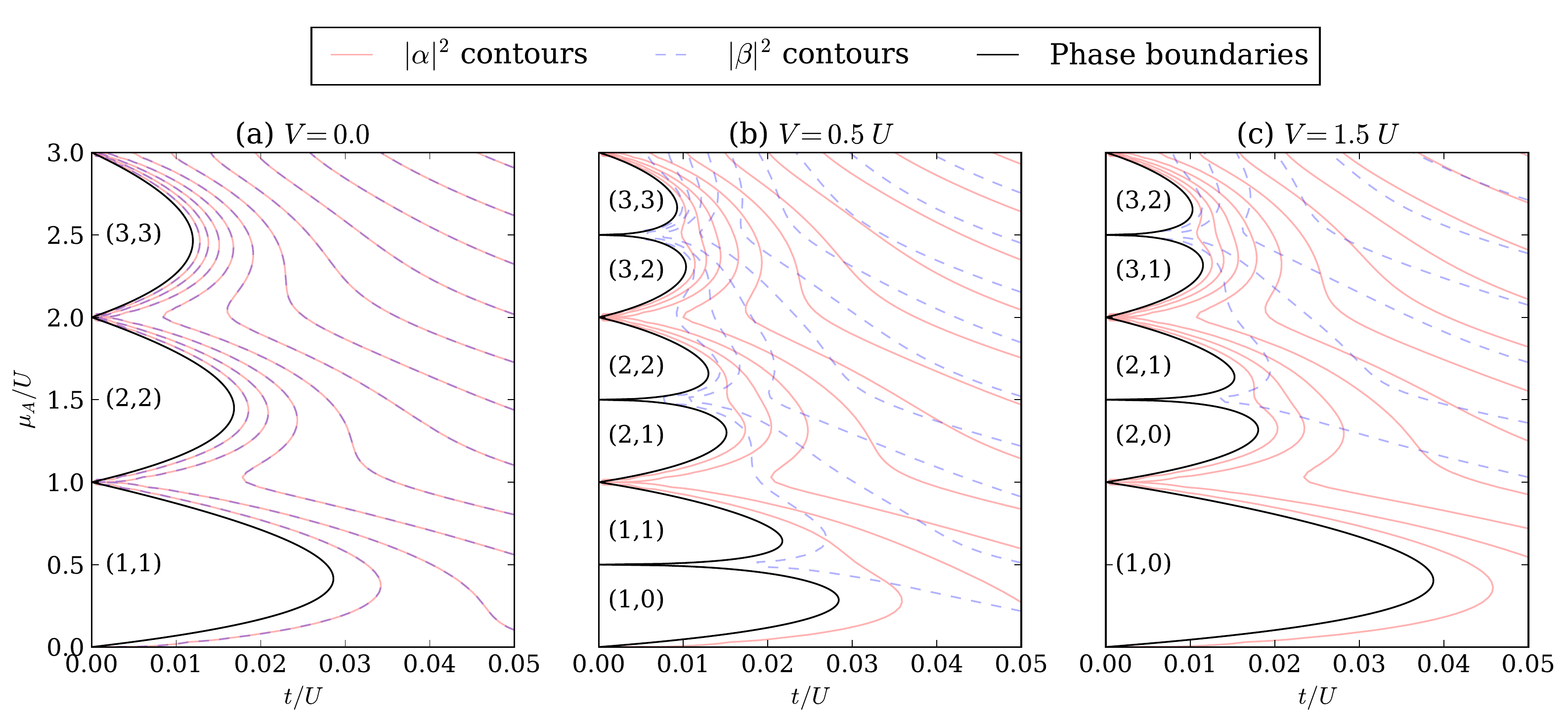}
  \caption{(color online) Phase boundaries for the kagome lattice for three different values of the sublattice bias $V$.  Energies are given in terms of the on-site interaction $U$.  Mott states with $n_A$ and $n_B$ atoms per site on each sublattice are labeled by $(n_A,n_B)$. Solid (red) curves and dashed (blue) curves denote evenly spaced contours of fixed condensate densities, $|\alpha|^2$ and $|\beta|^2$, with spacing 0.25.}
  \label{fig:lobes}
\end{figure*}

\section{Results}

As detailed in Appendix~\ref{mindetails}, we numerically minimize the energy as a function of the parameters in Eq.~(\ref{ham}).  In addition to superfluid phases, where $\alpha$ and $\beta$ are nonzero, we find insulator phases with $\alpha=\beta=0$.  These are incompressible phases with an integer number of particles $n_A,n_B$ on the sites of each sublattice.  Figure~\ref{fig:lobes} shows representative phase diagrams taking parameters corresponding to the kagome lattice shown in Fig.~\ref{fig:lattices}.  We take all hopping matrix elements equal to one-another, $t_A=t_B=t_{AB}=t$, but for this lattice there are no nearest neighbor sites on the $B$-sublattice, so $t_B$ does not matter.  Similarly, we take the interactions to be equal $U_A=U_B=U$.  We vary $\mu_A$ and the offset $V=\mu_B-\mu_A$.

\begin{figure}[tbhp]
  \centering
  \includegraphics[width=\linewidth]{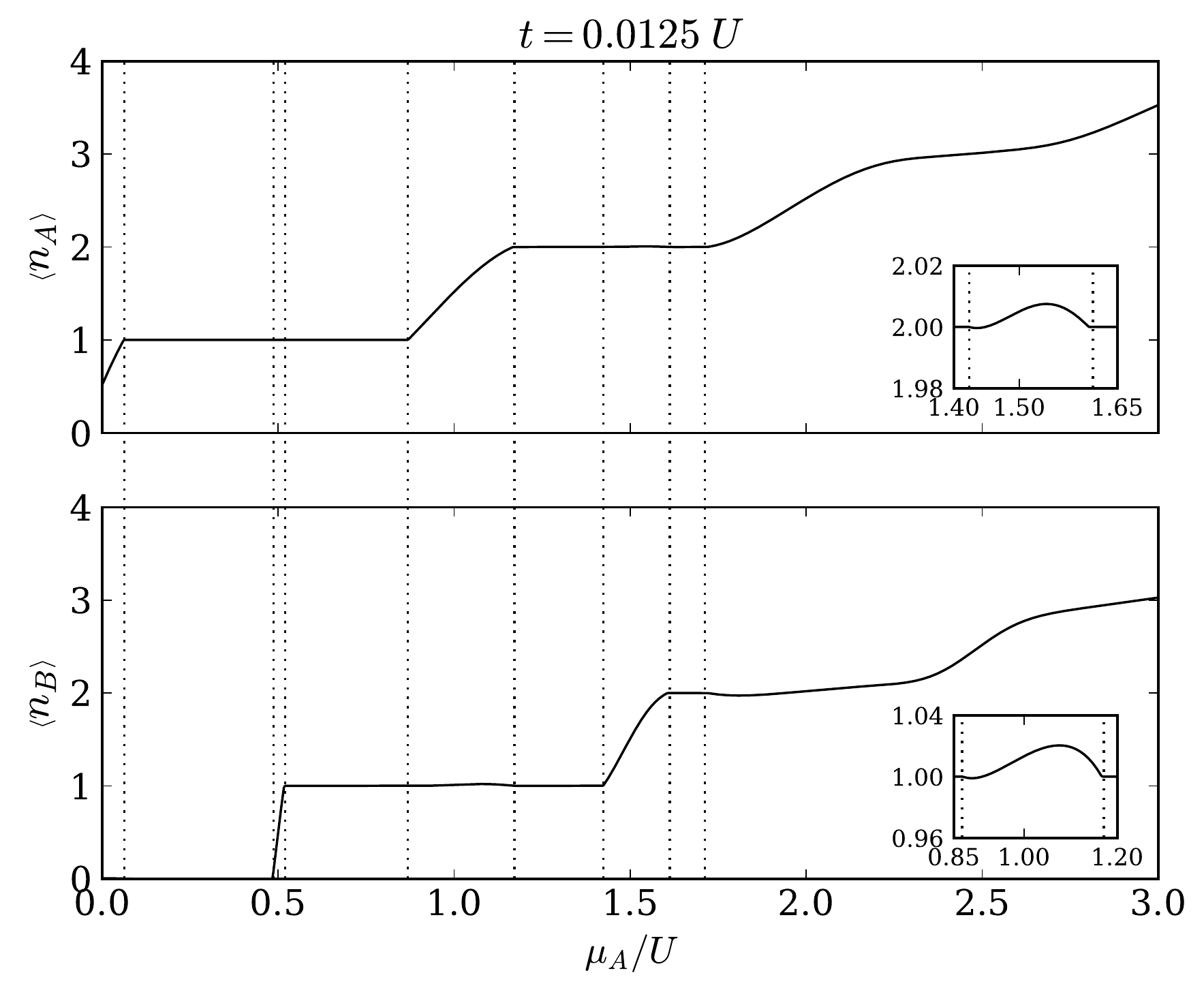}
  \caption{Densities on two sublattices of the kagome lattice for $V=0.5\;U$ and $t=0.0125\;U$ (see Fig.~\ref{fig:lobes}(b)).  Dotted vertical lines show phase boundaries. Insets show magnified views, with the same axes.}
  \label{fig:tslice}
\end{figure}

When $V=0$ this geometry reduces to a simple triangular lattice.  The Mott lobes have $n_A=n_B$, and the order parameters on the two sublattices are identical (Fig.~\ref{fig:lobes} (a)).  For small $V$ the lobes split: $n_A$ and $n_B$ increment independently (Fig.~\ref{fig:lobes} (b)).  
When $V>U$, there will be multiple lobes where $n_B=0$ (Fig.~\ref{fig:lobes}(c)).

In addition to showing phase boundaries, in Fig.~\ref{fig:lobes} we plot the surfaces of fixed condensate density, $|\alpha|^2$ and $|\beta|^2$.  As is evident in Fig.~\ref{fig:lobes}(b), there are regions where one order parameter is much bigger than the other.  These regions are found between pairs of lobes, and alternate:  when $n_A$ increments, one finds $|\alpha|^2\gg|\beta|^2$, and vice-versa when $n_B$ increments.  One interpretation of this asymmetry is that it is a manifestation of the proximity effect \cite{meissner}:  Between two lobes where $n_B$ is unchanged, the B-sublattice is nearly a Mott insulator, and it is only proximity to the A-superfluid which makes the order parameter non-zero.

\begin{figure}[tbhp]
  \centering
  \includegraphics[width=\linewidth]{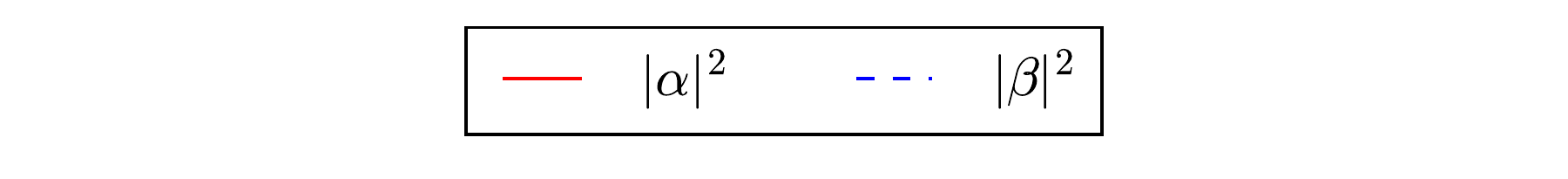}
  \includegraphics[width=\linewidth]{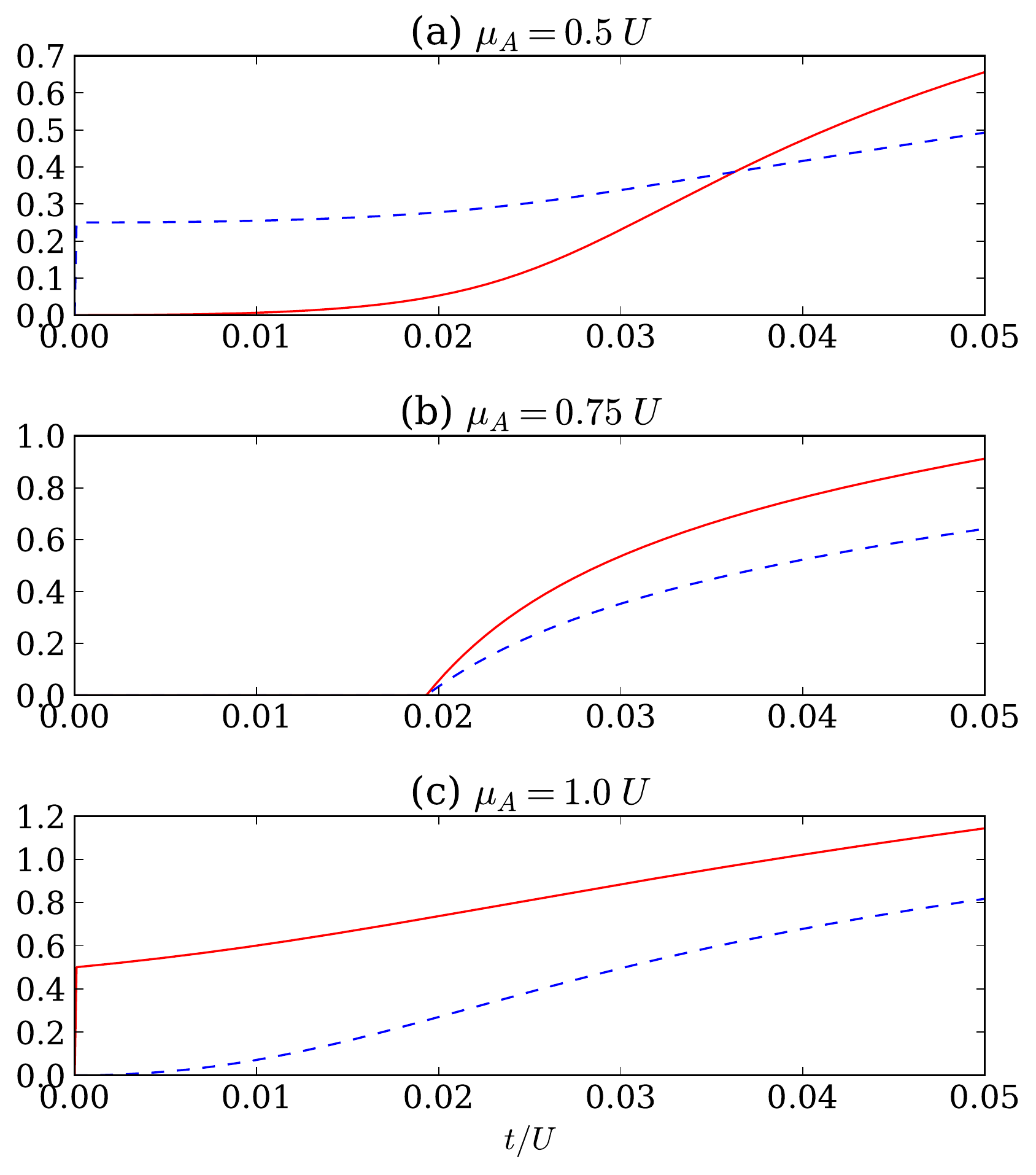}
  \caption{(color online)
  Order parameters on two sublattices of the kagome lattice for
  $V=0.5\;U$ and (a) $\mu_A=0.5\;U$, (b) $\mu_A=0.75\;U$, and (c) $\mu_A=U$.  Solid (red) and dashed (blue) curves correspond to $|\alpha|^2$ and $|\beta|^2$.}
  \label{fig:muslice}
\end{figure}

\begin{figure}[tbhp]
  \centering
  \includegraphics[width=\linewidth]{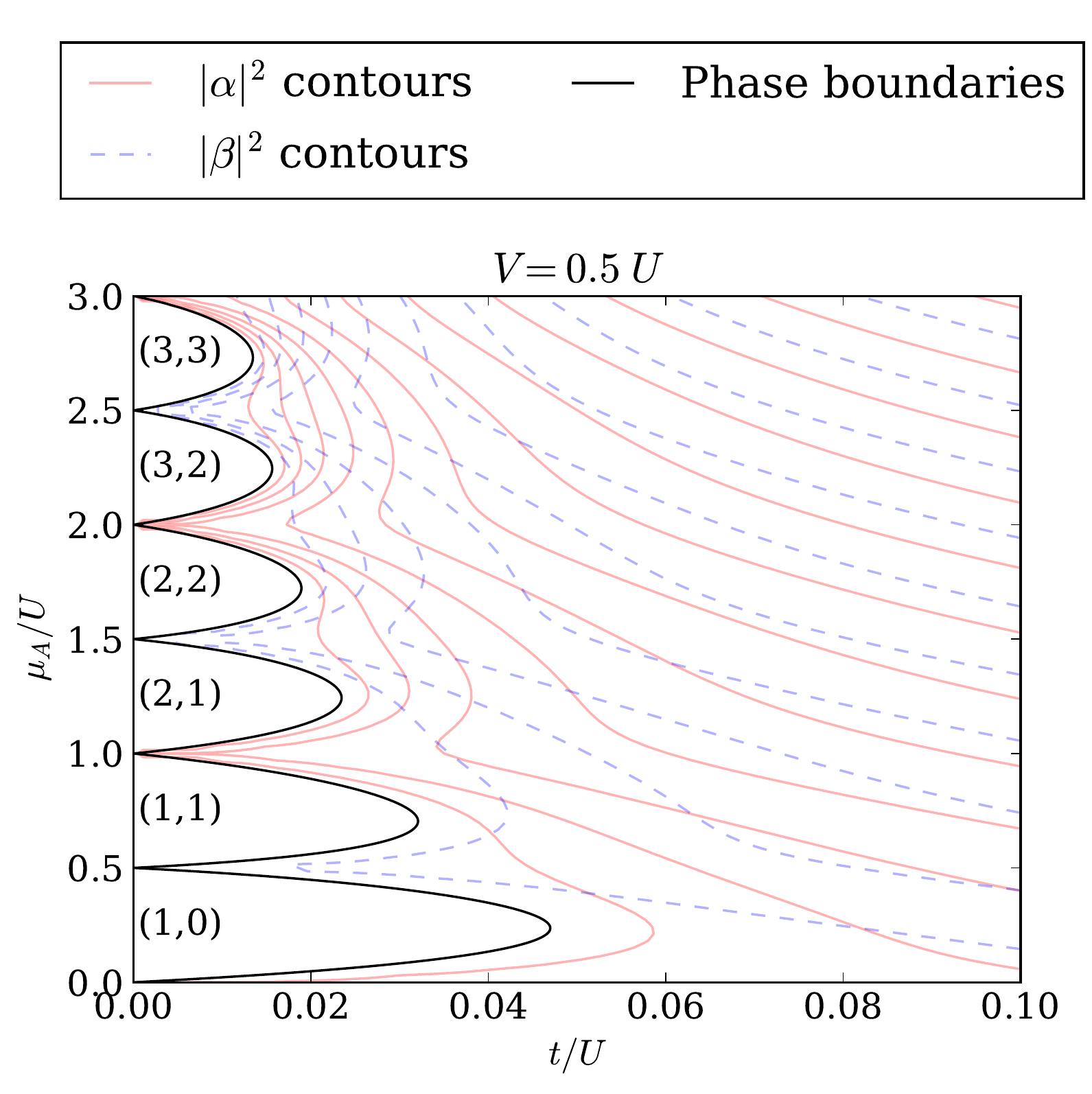}
  \caption{(color online) Phase boundaries for the striped lattice with sublattice bias $V=0.5$.  Energies are given in terms of the on-site interaction $U$.  Mott states with $n_A$ and $n_B$ atoms per site on each sublattice are labeled by $(n_A,n_B)$. Solid (red) curves and dashed (blue) curves denote evenly spaced contours of fixed condensate densities, $|\alpha|^2$ and $|\beta|^2$, with spacing 0.25. In contrast to the kagome lattice, for the striped lattice the lobes are more symmetric.}
  \label{fig:stripelobes}
\end{figure}

Another feature of Fig.~\ref{fig:lobes} is that the lobes where $n_A$ increment have larger gaps between them than those where $n_B$ increment.  This is due to the fact that there are no B-B neighbors, making the B superfluid less stable.  Figure~\ref{fig:stripelobes} shows the phase boundaries for the stripe lattice, where the two sublattices are equivalent.  There we see much more even spacing between the lobes.

In Fig.~\ref{fig:tslice}, we plot the densities as a function of $\mu_A$ for fixed $t$.  This corresponds to a vertical slice through the phase diagram.  One sees a series of plateaus, corresponding to the insulating states.  The two densities increment sequentially.  Notice that the density of one of the sublattices is nearly uniform in the superfluid region between two lobes.  Thus the fluid on that sublattice is approximately incompressible.  As apparent in the inset, if one zooms in, one can find some variation of the density in these regions.  Interestingly the slope $\partial\expval{n_\sigma}/\partial\mu_A$ can even be slightly negative, but the total compressibility $\partial\expval{n_A + n_B}/\partial\mu_A$ is always positive.

Figure~\ref{fig:muslice} illustrates the behavior of the order parameters as a function of $t/U$ for several fixed values of $\mu_A/U$.  This corresponds to horizontal slices through the phase diagram.  
Figure~\ref{fig:muslice}(a) shows a slice between lobes where $(n_a,n_b)=(1,0)$ and $(1,1)$.  The A-lattice order parameter becomes extremely small as $t\to0$, while the B-lattice order parameter approaches a constant.  Figure~\ref{fig:muslice}(c) shows the opposite case.  Figure~\ref{fig:muslice}(b) shows the generic case, where one has an insulator at small $t$.  For these parameters one is on the part of the lobe which is near the $(1,1)$ to $(2,1)$ transition, and hence $|\alpha|^2>|\beta|^2$.  In this generic case the condensate densities vanish linearly as one approaches the Mott lobe.

\section{Experimental detection}
A key question is how the sublattice superfluidity will manifest in an experiment.  The order parameters $|\alpha|^2$ and $|\beta|^2$ correspond to macroscopic occupation of $k=0$ states in each sublattice, and will therefore lead to sharp peaks in time-of-flight expansion \cite{tof}.  For the kagome lattice, the condensate contribution to the time-of-flight image is a series of peaks which again form a kagome lattice, with the dots and x's in Fig.~\ref{fig:lattices} having intensities proportional to $|3 \alpha+\beta|^2$ and $|\alpha-\beta|^2$.  Thus by comparing the intensities of the Bragg peaks, one can extract the ratios of the condensate fractions.  The full argument is given in Appendix~\ref{sec:tof}.

An alternative probe is site-resolved in-situ imaging \cite{insitu}, which can measure the densities $n_A$ and $n_B$.  In principle one could thereby identify the Mott states.  As illustrated in Fig.~\ref{fig:tslice}, there are however regions of parameter space in which the superfluid has nearly an integer number of particles per site.  These regions may make it difficult to use in-situ imaging to find the phase boundaries.

Another option for detecting the density differences between sublattices is light scattering. This technique has been used to find antiferromagnetic spin correlations \cite{lightscattering}.

\section{Summary}
We have studied the influence of sublattice bias on the superfluid to insulator transition of a Bose gas in an optical lattice. Within the Gutzwiller mean field theory, superfluidity on one sublattice is always accompanied by superfluidity on the other. The Mott lobes are characterized by integer densities on each sublattice. These increment sequentially as the chemical potential is increased. Between the lobes there are regions where one superfluid density is much larger than the other. These features are observable through in-situ and time-of-flight probes.

Although beyond the scope of this study, these biased lattices may be a promising platform for searching for more exotic physics. For example, related geometries have been used to study fractional Mott insulators \cite{fractionalmott1,fractionalmott2}.

\section{Acknowledgements}

This material is based upon work supported by the National Science Foundation under Grant No. PHY-1508300.

\appendix

\section{Minimizing the energy}
\label{mindetails}
Within our variational theory, the energy $\langle \mathcal{H}\rangle$ is a quartic form in the coefficients $f^\sigma_n$ and their complex conjugates, where $\sigma=A,B$ labels the sublattice.  The quartic terms are products of the mean-fields $\psi^A\equiv\alpha$ and $\psi^B\equiv\beta$, and their complex conjugates.  This feature, combined with the relation $\partial \psi^\sigma/\partial (f_n^\sigma)^*=\sqrt{n} f^\sigma_{n-1}$ means that the derivative $\partial \langle \mathcal{H}\rangle/\partial (f^\sigma_n)^*$ has a simple form.  Making $\langle H\rangle$ stationary, with the constraint $\bra{\psi}\ket{\psi} = 1$, yields a set of cubic equations, that can be written
 \begin{gather}
    \Lambda_n^\sigma f_n^\sigma + \Gamma_{n+1}^\sigma f_{n+1}^\sigma + \Gamma_n^{\sigma^*} f_{n-1}^\sigma = \lambda^\sigma f_n^\sigma
    \label{eqn:numerical}
\end{gather}
where $\sigma, \tau \in \{A, B\}$, and 
\begin{align}\label{coefs}
    \Lambda_n^\sigma &= \frac{N_\sigma}{2} n(n-1) - N_\sigma \frac{\mu_\sigma}{U} n\\
    \Gamma_n^\sigma &= \sum_{\tau} \left(-\frac{t_{\sigma\tau}}{U}\right)(N_\sigma s_{\sigma\tau} \sqrt{n}\; \psi^\tau)
\end{align}
Here $N_\sigma$ is the number of sites in sublattice $\sigma$  and $s_{\sigma\tau}$ is the number of $\tau$ neighbors of each $\sigma$ site. $\Lambda$ is independent of the $f$'s, while $\Gamma$ depends on the $f$'s only through $\psi$.

The cubic eigenvalue equation in Eq.~(\ref{eqn:numerical}) can be solved efficiently by an iterative method.  We start with a guess for $\psi^\tau$ in Eq.~(\ref{coefs}). We then solve Eq.~(\ref{eqn:numerical}) as a linear eigenvalue equation.  That solution is then used to update $\psi^\tau$.  Both the time for each iteration, and the total number of iterations needed for convergence, are favorable.

\section{Phase boundaries}
\label{sec:phasebound}
The Mott insulator regions in the $\frac{\mu}{U}$-$\frac{t}{U}$ parameter space can be labelled by integers $n_A = \expval{a_i^\dagger a_i}$ and  $n_B = \expval{b_i^\dagger b_i}$. Consider a point in the superfluid region infinitesimally close to the insulator region with $n_A = j, n_B = k$. To the leading order, we must have
\begin{align*}
\begin{aligned}[c]
f^A_{j-1} &= \epsilon_- \\
f^A_{j+1} &= \epsilon_+ \\
f^A_{j} &= \sqrt{1 - \epsilon_-^2 - \epsilon_+^2} \\
\end{aligned}
\qquad
\begin{aligned}[c]
f^B_{k-1} &= \eta_- \\
f^B_{k+1} &= \eta_+ \\
f^B_{k} &= \sqrt{1 - \eta_-^2 - \eta_+^2} \\
\end{aligned}
\end{align*}
All the other $f$'s vanish. As we approach the insulator, the parameters $\{\epsilon, \eta\}$ become smaller, vanishing at the phase boundary.
Consequently, we can expand the energy as
\begin{align*}
\expval{H} = \left(\begin{array}{cccc} \epsilon_+ & \epsilon_- & \eta_+ & \eta_- \end{array}\right)\cdot M \cdot \left(\begin{array}{c} \epsilon_+ \\ \epsilon_- \\ \eta_+ \\ \eta_- \end{array}\right) + \cdots
\end{align*}
where $M$ is a $4\times4$ matrix and in addition to a constant, the neglected terms contain higher powers of the small parameters. The phase transition is characterized by $\det(M) = 0$, as $M$ is positive definite in the insulating phase, but has at least one negative eigenvalue in the superfluid phase. This equation gives the phase transition boundary in the $\frac{\mu}{U}$-$\frac{t}{U}$ parameter space.

The equation $\det(M)=0$ can be factored as the product of two quadratic equations in $t$ which have closed form solutions. These analytic expressions are not particularly enlightening.

\section{Time of flight}
\label{sec:tof}

As explained in the main text, the time-of-flight density pattern can reveal the sublattice condensate fraction. Here we calculate this pattern for a kagome lattice with sublattice bias.

The time-of-flight image from a lattice gas contains two components: a diffuse background from the non-condensed particles, and sharp Bragg peaks from the condensate. Here we solely consider the latter, which are readily distinguished. As explained in \cite{tof}, the asymptotic profile is simply given by the Fourier transform of the in-situ condensate wavefunction, $\psi(\mathbf{k}, t) \propto \psi_{\mathbf{k}=\frac{m\mathbf{r}}{t}}$ where $\psi_\mathbf{k} = \int \mathop{d\mathbf{r}} e^{-i\mathbf{k}\cdot\mathbf{r}} \psi(\mathbf{r}, 0)$.

The initial state of the system can be written as
\begin{align}
\psi(\mathbf{r}, 0) &= \sum_{i\in A} \alpha \phi(\mathbf{r} - \mathbf{r_i}) + \sum_{i\in B}\beta \phi(\mathbf{r}-\mathbf{r_i})
\end{align}
with the Fourier transform,
\begin{align}
\psi_\mathbf{k}(0) = \sum_{i\in A} \alpha e^{-i\mathbf{k}\cdot\mathbf{r}_i} \phi_\mathbf{k} + \sum_{i\in B}\beta e^{-i\mathbf{k}\cdot\mathbf{r}_i} \phi_\mathbf{k}
\end{align}
where $\phi_\mathbf{k}$ is the Fourier transform of the Wannier state $\phi(\mathbf{r})$.

The sites in sublattice $A$ belong to one of three Bravais lattices $\mathbf{r}_s = m\mathbf{u} + n\mathbf{v} + \mathbf{\delta}_s$, where $\mathbf{u} = (a, 0)$, $\mathbf{v} = (a\cos \frac{\pi}{3}, a\sin \frac{\pi}{3})$, $m$ and $n$ are integers, and $\mathbf{\delta}_s$ takes on one of three values: $\mathbf{\delta}_0 = (0,0)$, $\mathbf{\delta}_1 = \mathbf{u}/2$, $\mathbf{\delta}_2 = \mathbf{v}/2$. The sites in sublattice $B$ belong to the Bravais lattice $\mathbf{r}_3 = m\mathbf{u} + n\mathbf{v} + \mathbf{\delta}_3$ with $\mathbf{\delta}_3 = (\mathbf{u} + \mathbf{v})/2$. Then, we can write
\begin{equation}
\begin{aligned}
\label{eqn:tof}
\sum_{i\in A} e^{-i\mathbf{k}\cdot\mathbf{r}_i} &= F(\mathbf{k})\,\left(1 + e^{-i\mathbf{k}\cdot\frac{\mathbf{u}}{2}} + e^{-i\mathbf{k}\cdot\frac{\mathbf{v}}{2}}\right)\\
\sum_{i\in B} e^{-i\mathbf{k}\cdot\mathbf{r}_i} &= F(\mathbf{k})\, e^{-i\mathbf{k}\cdot\left(\frac{\mathbf{u}}{2} + \frac{\mathbf{v}}{2}\right)}
\end{aligned}
\end{equation}
where $F(\mathbf{k}) = \sum_{m,n} e^{-i\mathbf{k}\cdot(m\mathbf{u} + n\mathbf{v})}$ is zero unless $\mathbf{k}\cdot(m\mathbf{u} + n\mathbf{v})$ is a multiple of $2\pi$ for all integer $m,n$.  We define reciprocal lattice vectos $\mathbf{k}_1, \mathbf{k}_2$ such that $\mathbf{k}_1\cdot\mathbf{u}=2\pi$, $\mathbf{k}_1\cdot\mathbf{v}=0$, $\mathbf{k}_2\cdot\mathbf{u}=0$, $\mathbf{k}_2\cdot\mathbf{v}=2\pi$. In particular, $\mathbf{k}_1 = \left(\frac{2\pi}{a}, -\frac{2\pi}{a}\cot\frac{\pi}{3}\right)$, $\mathbf{k}_2 = \left(0, \frac{2\pi}{a}\csc\frac{\pi}{3}\right)$ are the basis for a triangular lattice. Standard analysis gives $F(\mathbf{k}) \propto \sum_{m,n} \delta(\mathbf{k} - m\mathbf{k}_1 - n\mathbf{k}_2)$. 

Because of coherence factors in Eq.~(\ref{eqn:tof}), the number of atoms in the peak at position $\mathbf{r}_{mn} = \frac{t}{m}(m\mathbf{k}_1, n\mathbf{k}_2)$ is
\begin{align}
    N_{nm} \propto \abs{\alpha(1 + (-1)^n + (-1)^m) + \beta(-1)^{n+m}}^2 \abs{\phi_{nm}}^2
\end{align}
where $\phi_{nm} = \phi_{\mathbf{k}}$ at $\mathbf{k} = m\mathbf{r}/t$. Neglecting this slow envelope yields
\begin{align}
    N_{00} = N_{22} &\propto \abs{3\alpha + \beta}^2 \notag\\
    N_{11} = N_{10} = N_{01} &\propto\abs{\alpha - \beta}^2
\end{align}

\bibliographystyle{apsrev4-1}
\bibliography{references}

\begin{thebibliography}{31}%
\makeatletter
\providecommand \@ifxundefined [1]{%
 \@ifx{#1\undefined}
}%
\providecommand \@ifnum [1]{%
 \ifnum #1\expandafter \@firstoftwo
 \else \expandafter \@secondoftwo
 \fi
}%
\providecommand \@ifx [1]{%
 \ifx #1\expandafter \@firstoftwo
 \else \expandafter \@secondoftwo
 \fi
}%
\providecommand \natexlab [1]{#1}%
\providecommand \enquote  [1]{``#1''}%
\providecommand \bibnamefont  [1]{#1}%
\providecommand \bibfnamefont [1]{#1}%
\providecommand \citenamefont [1]{#1}%
\providecommand \href@noop [0]{\@secondoftwo}%
\providecommand \href [0]{\begingroup \@sanitize@url \@href}%
\providecommand \@href[1]{\@@startlink{#1}\@@href}%
\providecommand \@@href[1]{\endgroup#1\@@endlink}%
\providecommand \@sanitize@url [0]{\catcode `\\12\catcode `\$12\catcode
  `\&12\catcode `\#12\catcode `\^12\catcode `\_12\catcode `\%12\relax}%
\providecommand \@@startlink[1]{}%
\providecommand \@@endlink[0]{}%
\providecommand \url  [0]{\begingroup\@sanitize@url \@url }%
\providecommand \@url [1]{\endgroup\@href {#1}{\urlprefix }}%
\providecommand \urlprefix  [0]{URL }%
\providecommand \Eprint [0]{\href }%
\providecommand \doibase [0]{http://dx.doi.org/}%
\providecommand \selectlanguage [0]{\@gobble}%
\providecommand \bibinfo  [0]{\@secondoftwo}%
\providecommand \bibfield  [0]{\@secondoftwo}%
\providecommand \translation [1]{[#1]}%
\providecommand \BibitemOpen [0]{}%
\providecommand \bibitemStop [0]{}%
\providecommand \bibitemNoStop [0]{.\EOS\space}%
\providecommand \EOS [0]{\spacefactor3000\relax}%
\providecommand \BibitemShut  [1]{\csname bibitem#1\endcsname}%
\let\auto@bib@innerbib\@empty
\bibitem [{\citenamefont {Greiner}\ \emph {et~al.}(2002)\citenamefont
  {Greiner}, \citenamefont {Mandel}, \citenamefont {Esslinger}, \citenamefont
  {H{\"a}nsch},\ and\ \citenamefont {Bloch}}]{greiner}%
  \BibitemOpen
  \bibfield  {author} {\bibinfo {author} {\bibfnamefont {M.}~\bibnamefont
  {Greiner}}, \bibinfo {author} {\bibfnamefont {O.}~\bibnamefont {Mandel}},
  \bibinfo {author} {\bibfnamefont {T.}~\bibnamefont {Esslinger}}, \bibinfo
  {author} {\bibfnamefont {T.~W.}\ \bibnamefont {H{\"a}nsch}}, \ and\ \bibinfo
  {author} {\bibfnamefont {I.}~\bibnamefont {Bloch}},\ }\href {\doibase
  10.1038/415039a} {\bibfield  {journal} {\bibinfo  {journal} {Nature}\
  }\textbf {\bibinfo {volume} {415}},\ \bibinfo {pages} {39} (\bibinfo {year}
  {2002})}\BibitemShut {NoStop}%
\bibitem [{\citenamefont {DeMarco}\ \emph {et~al.}(2005)\citenamefont
  {DeMarco}, \citenamefont {Lannert}, \citenamefont {Vishveshwara},\ and\
  \citenamefont {Wei}}]{demarco2005}%
  \BibitemOpen
  \bibfield  {author} {\bibinfo {author} {\bibfnamefont {B.}~\bibnamefont
  {DeMarco}}, \bibinfo {author} {\bibfnamefont {C.}~\bibnamefont {Lannert}},
  \bibinfo {author} {\bibfnamefont {S.}~\bibnamefont {Vishveshwara}}, \ and\
  \bibinfo {author} {\bibfnamefont {T.-C.}\ \bibnamefont {Wei}},\ }\href
  {\doibase 10.1103/physreva.71.063601} {\bibfield  {journal} {\bibinfo
  {journal} {Physical Review A}\ }\textbf {\bibinfo {volume} {71}} (\bibinfo
  {year} {2005}),\ 10.1103/physreva.71.063601}\BibitemShut {NoStop}%
\bibitem [{\citenamefont {Campbell}(2006)}]{campbell2006}%
  \BibitemOpen
  \bibfield  {author} {\bibinfo {author} {\bibfnamefont {G.~K.}\ \bibnamefont
  {Campbell}},\ }\href {\doibase 10.1126/science.1130365} {\bibfield  {journal}
  {\bibinfo  {journal} {Science}\ }\textbf {\bibinfo {volume} {313}},\ \bibinfo
  {pages} {649} (\bibinfo {year} {2006})}\BibitemShut {NoStop}%
\bibitem [{\citenamefont {F\"{o}lling}\ \emph {et~al.}(2006)\citenamefont
  {F\"{o}lling}, \citenamefont {Widera}, \citenamefont {M\"{u}ller},
  \citenamefont {Gerbier},\ and\ \citenamefont {Bloch}}]{folling2006}%
  \BibitemOpen
  \bibfield  {author} {\bibinfo {author} {\bibfnamefont {S.}~\bibnamefont
  {F\"{o}lling}}, \bibinfo {author} {\bibfnamefont {A.}~\bibnamefont {Widera}},
  \bibinfo {author} {\bibfnamefont {T.}~\bibnamefont {M\"{u}ller}}, \bibinfo
  {author} {\bibfnamefont {F.}~\bibnamefont {Gerbier}}, \ and\ \bibinfo
  {author} {\bibfnamefont {I.}~\bibnamefont {Bloch}},\ }\href {\doibase
  10.1103/physrevlett.97.060403} {\bibfield  {journal} {\bibinfo  {journal}
  {Physical Review Letters}\ }\textbf {\bibinfo {volume} {97}} (\bibinfo {year}
  {2006}),\ 10.1103/physrevlett.97.060403}\BibitemShut {NoStop}%
\bibitem [{\citenamefont {Gerbier}\ \emph
  {et~al.}(2005{\natexlab{a}})\citenamefont {Gerbier}, \citenamefont {Widera},
  \citenamefont {F\"{o}lling}, \citenamefont {Mandel}, \citenamefont
  {Gericke},\ and\ \citenamefont {Bloch}}]{gerbier2005a}%
  \BibitemOpen
  \bibfield  {author} {\bibinfo {author} {\bibfnamefont {F.}~\bibnamefont
  {Gerbier}}, \bibinfo {author} {\bibfnamefont {A.}~\bibnamefont {Widera}},
  \bibinfo {author} {\bibfnamefont {S.}~\bibnamefont {F\"{o}lling}}, \bibinfo
  {author} {\bibfnamefont {O.}~\bibnamefont {Mandel}}, \bibinfo {author}
  {\bibfnamefont {T.}~\bibnamefont {Gericke}}, \ and\ \bibinfo {author}
  {\bibfnamefont {I.}~\bibnamefont {Bloch}},\ }\href {\doibase
  10.1103/physreva.72.053606} {\bibfield  {journal} {\bibinfo  {journal}
  {Physical Review A}\ }\textbf {\bibinfo {volume} {72}} (\bibinfo {year}
  {2005}{\natexlab{a}}),\ 10.1103/physreva.72.053606}\BibitemShut {NoStop}%
\bibitem [{\citenamefont {Gerbier}\ \emph
  {et~al.}(2005{\natexlab{b}})\citenamefont {Gerbier}, \citenamefont {Widera},
  \citenamefont {F\"{o}lling}, \citenamefont {Mandel}, \citenamefont
  {Gericke},\ and\ \citenamefont {Bloch}}]{gerbier2005b}%
  \BibitemOpen
  \bibfield  {author} {\bibinfo {author} {\bibfnamefont {F.}~\bibnamefont
  {Gerbier}}, \bibinfo {author} {\bibfnamefont {A.}~\bibnamefont {Widera}},
  \bibinfo {author} {\bibfnamefont {S.}~\bibnamefont {F\"{o}lling}}, \bibinfo
  {author} {\bibfnamefont {O.}~\bibnamefont {Mandel}}, \bibinfo {author}
  {\bibfnamefont {T.}~\bibnamefont {Gericke}}, \ and\ \bibinfo {author}
  {\bibfnamefont {I.}~\bibnamefont {Bloch}},\ }\href {\doibase
  10.1103/physrevlett.95.050404} {\bibfield  {journal} {\bibinfo  {journal}
  {Physical Review Letters}\ }\textbf {\bibinfo {volume} {95}} (\bibinfo {year}
  {2005}{\natexlab{b}}),\ 10.1103/physrevlett.95.050404}\BibitemShut {NoStop}%
\bibitem [{\citenamefont {Orzel}(2001)}]{orzel2001}%
  \BibitemOpen
  \bibfield  {author} {\bibinfo {author} {\bibfnamefont {C.}~\bibnamefont
  {Orzel}},\ }\href {\doibase 10.1126/science.1058149} {\bibfield  {journal}
  {\bibinfo  {journal} {Science}\ }\textbf {\bibinfo {volume} {291}},\ \bibinfo
  {pages} {2386} (\bibinfo {year} {2001})}\BibitemShut {NoStop}%
\bibitem [{\citenamefont {Hadzibabic}\ \emph {et~al.}(2004)\citenamefont
  {Hadzibabic}, \citenamefont {Stock}, \citenamefont {Battelier}, \citenamefont
  {Bretin},\ and\ \citenamefont {Dalibard}}]{hadzibabic2004}%
  \BibitemOpen
  \bibfield  {author} {\bibinfo {author} {\bibfnamefont {Z.}~\bibnamefont
  {Hadzibabic}}, \bibinfo {author} {\bibfnamefont {S.}~\bibnamefont {Stock}},
  \bibinfo {author} {\bibfnamefont {B.}~\bibnamefont {Battelier}}, \bibinfo
  {author} {\bibfnamefont {V.}~\bibnamefont {Bretin}}, \ and\ \bibinfo {author}
  {\bibfnamefont {J.}~\bibnamefont {Dalibard}},\ }\href {\doibase
  10.1103/physrevlett.93.180403} {\bibfield  {journal} {\bibinfo  {journal}
  {Physical Review Letters}\ }\textbf {\bibinfo {volume} {93}} (\bibinfo {year}
  {2004}),\ 10.1103/physrevlett.93.180403}\BibitemShut {NoStop}%
\bibitem [{\citenamefont {Schori}\ \emph {et~al.}(2004)\citenamefont {Schori},
  \citenamefont {St\"{o}ferle}, \citenamefont {Moritz}, \citenamefont
  {K\"{o}hl},\ and\ \citenamefont {Esslinger}}]{schori2004}%
  \BibitemOpen
  \bibfield  {author} {\bibinfo {author} {\bibfnamefont {C.}~\bibnamefont
  {Schori}}, \bibinfo {author} {\bibfnamefont {T.}~\bibnamefont
  {St\"{o}ferle}}, \bibinfo {author} {\bibfnamefont {H.}~\bibnamefont
  {Moritz}}, \bibinfo {author} {\bibfnamefont {M.}~\bibnamefont {K\"{o}hl}}, \
  and\ \bibinfo {author} {\bibfnamefont {T.}~\bibnamefont {Esslinger}},\ }\href
  {\doibase 10.1103/physrevlett.93.240402} {\bibfield  {journal} {\bibinfo
  {journal} {Physical Review Letters}\ }\textbf {\bibinfo {volume} {93}}
  (\bibinfo {year} {2004}),\ 10.1103/physrevlett.93.240402}\BibitemShut
  {NoStop}%
\bibitem [{\citenamefont {Xu}\ \emph {et~al.}(2006)\citenamefont {Xu},
  \citenamefont {Liu}, \citenamefont {Miller}, \citenamefont {Chin},
  \citenamefont {Setiawan},\ and\ \citenamefont {Ketterle}}]{xu2006}%
  \BibitemOpen
  \bibfield  {author} {\bibinfo {author} {\bibfnamefont {K.}~\bibnamefont
  {Xu}}, \bibinfo {author} {\bibfnamefont {Y.}~\bibnamefont {Liu}}, \bibinfo
  {author} {\bibfnamefont {D.~E.}\ \bibnamefont {Miller}}, \bibinfo {author}
  {\bibfnamefont {J.~K.}\ \bibnamefont {Chin}}, \bibinfo {author}
  {\bibfnamefont {W.}~\bibnamefont {Setiawan}}, \ and\ \bibinfo {author}
  {\bibfnamefont {W.}~\bibnamefont {Ketterle}},\ }\href {\doibase
  10.1103/physrevlett.96.180405} {\bibfield  {journal} {\bibinfo  {journal}
  {Physical Review Letters}\ }\textbf {\bibinfo {volume} {96}} (\bibinfo {year}
  {2006}),\ 10.1103/physrevlett.96.180405}\BibitemShut {NoStop}%
\bibitem [{\citenamefont {Sengupta}\ \emph {et~al.}(2005)\citenamefont
  {Sengupta}, \citenamefont {Rigol}, \citenamefont {Batrouni}, \citenamefont
  {Denteneer},\ and\ \citenamefont {Scalettar}}]{sengupta2005}%
  \BibitemOpen
  \bibfield  {author} {\bibinfo {author} {\bibfnamefont {P.}~\bibnamefont
  {Sengupta}}, \bibinfo {author} {\bibfnamefont {M.}~\bibnamefont {Rigol}},
  \bibinfo {author} {\bibfnamefont {G.~G.}\ \bibnamefont {Batrouni}}, \bibinfo
  {author} {\bibfnamefont {P.~J.~H.}\ \bibnamefont {Denteneer}}, \ and\
  \bibinfo {author} {\bibfnamefont {R.~T.}\ \bibnamefont {Scalettar}},\ }\href
  {\doibase 10.1103/physrevlett.95.220402} {\bibfield  {journal} {\bibinfo
  {journal} {Physical Review Letters}\ }\textbf {\bibinfo {volume} {95}}
  (\bibinfo {year} {2005}),\ 10.1103/physrevlett.95.220402}\BibitemShut
  {NoStop}%
\bibitem [{\citenamefont {St\"{o}ferle}\ \emph {et~al.}(2004)\citenamefont
  {St\"{o}ferle}, \citenamefont {Moritz}, \citenamefont {Schori}, \citenamefont
  {K\"{o}hl},\ and\ \citenamefont {Esslinger}}]{stoferle2004}%
  \BibitemOpen
  \bibfield  {author} {\bibinfo {author} {\bibfnamefont {T.}~\bibnamefont
  {St\"{o}ferle}}, \bibinfo {author} {\bibfnamefont {H.}~\bibnamefont
  {Moritz}}, \bibinfo {author} {\bibfnamefont {C.}~\bibnamefont {Schori}},
  \bibinfo {author} {\bibfnamefont {M.}~\bibnamefont {K\"{o}hl}}, \ and\
  \bibinfo {author} {\bibfnamefont {T.}~\bibnamefont {Esslinger}},\ }\href
  {\doibase 10.1103/physrevlett.92.130403} {\bibfield  {journal} {\bibinfo
  {journal} {Physical Review Letters}\ }\textbf {\bibinfo {volume} {92}}
  (\bibinfo {year} {2004}),\ 10.1103/physrevlett.92.130403}\BibitemShut
  {NoStop}%
\bibitem [{\citenamefont {Gerbier}\ \emph {et~al.}(2006)\citenamefont
  {Gerbier}, \citenamefont {F\"{o}lling}, \citenamefont {Widera}, \citenamefont
  {Mandel},\ and\ \citenamefont {Bloch}}]{gerbier2006}%
  \BibitemOpen
  \bibfield  {author} {\bibinfo {author} {\bibfnamefont {F.}~\bibnamefont
  {Gerbier}}, \bibinfo {author} {\bibfnamefont {S.}~\bibnamefont
  {F\"{o}lling}}, \bibinfo {author} {\bibfnamefont {A.}~\bibnamefont {Widera}},
  \bibinfo {author} {\bibfnamefont {O.}~\bibnamefont {Mandel}}, \ and\ \bibinfo
  {author} {\bibfnamefont {I.}~\bibnamefont {Bloch}},\ }\href {\doibase
  10.1103/physrevlett.96.090401} {\bibfield  {journal} {\bibinfo  {journal}
  {Physical Review Letters}\ }\textbf {\bibinfo {volume} {96}} (\bibinfo {year}
  {2006}),\ 10.1103/physrevlett.96.090401}\BibitemShut {NoStop}%
\bibitem [{\citenamefont {Bakr}\ \emph {et~al.}(2010)\citenamefont {Bakr},
  \citenamefont {Peng}, \citenamefont {Tai}, \citenamefont {Ma}, \citenamefont
  {Simon}, \citenamefont {Gillen}, \citenamefont {Folling}, \citenamefont
  {Pollet},\ and\ \citenamefont {Greiner}}]{greiner2010}%
  \BibitemOpen
  \bibfield  {author} {\bibinfo {author} {\bibfnamefont {W.~S.}\ \bibnamefont
  {Bakr}}, \bibinfo {author} {\bibfnamefont {A.}~\bibnamefont {Peng}}, \bibinfo
  {author} {\bibfnamefont {M.~E.}\ \bibnamefont {Tai}}, \bibinfo {author}
  {\bibfnamefont {R.}~\bibnamefont {Ma}}, \bibinfo {author} {\bibfnamefont
  {J.}~\bibnamefont {Simon}}, \bibinfo {author} {\bibfnamefont {J.~I.}\
  \bibnamefont {Gillen}}, \bibinfo {author} {\bibfnamefont {S.}~\bibnamefont
  {Folling}}, \bibinfo {author} {\bibfnamefont {L.}~\bibnamefont {Pollet}}, \
  and\ \bibinfo {author} {\bibfnamefont {M.}~\bibnamefont {Greiner}},\ }\href
  {\doibase 10.1126/science.1192368} {\bibfield  {journal} {\bibinfo  {journal}
  {Science}\ }\textbf {\bibinfo {volume} {329}},\ \bibinfo {pages} {547}
  (\bibinfo {year} {2010})}\BibitemShut {NoStop}%
\bibitem [{\citenamefont {Jo}\ \emph {et~al.}(2012)\citenamefont {Jo},
  \citenamefont {Guzman}, \citenamefont {Thomas}, \citenamefont {Hosur},
  \citenamefont {Vishwanath},\ and\ \citenamefont
  {Stamper-Kurn}}]{stamperkurn}%
  \BibitemOpen
  \bibfield  {author} {\bibinfo {author} {\bibfnamefont {G.-B.}\ \bibnamefont
  {Jo}}, \bibinfo {author} {\bibfnamefont {J.}~\bibnamefont {Guzman}}, \bibinfo
  {author} {\bibfnamefont {C.~K.}\ \bibnamefont {Thomas}}, \bibinfo {author}
  {\bibfnamefont {P.}~\bibnamefont {Hosur}}, \bibinfo {author} {\bibfnamefont
  {A.}~\bibnamefont {Vishwanath}}, \ and\ \bibinfo {author} {\bibfnamefont
  {D.~M.}\ \bibnamefont {Stamper-Kurn}},\ }\href {\doibase
  10.1103/physrevlett.108.045305} {\bibfield  {journal} {\bibinfo  {journal}
  {Physical Review Letters}\ }\textbf {\bibinfo {volume} {108}} (\bibinfo
  {year} {2012}),\ 10.1103/physrevlett.108.045305}\BibitemShut {NoStop}%
\bibitem [{\citenamefont {Natu}\ \emph {et~al.}(2016)\citenamefont {Natu},
  \citenamefont {Mueller},\ and\ \citenamefont {Sarma}}]{Natu2016}%
  \BibitemOpen
  \bibfield  {author} {\bibinfo {author} {\bibfnamefont {S.~S.}\ \bibnamefont
  {Natu}}, \bibinfo {author} {\bibfnamefont {E.~J.}\ \bibnamefont {Mueller}}, \
  and\ \bibinfo {author} {\bibfnamefont {S.~D.}\ \bibnamefont {Sarma}},\ }\href
  {\doibase 10.1103/physreva.93.063610} {\bibfield  {journal} {\bibinfo
  {journal} {Physical Review A}\ }\textbf {\bibinfo {volume} {93}} (\bibinfo
  {year} {2016}),\ 10.1103/physreva.93.063610}\BibitemShut {NoStop}%
\bibitem [{\citenamefont {Goldbaum}\ and\ \citenamefont
  {Mueller}(2008)}]{Goldbaum2008}%
  \BibitemOpen
  \bibfield  {author} {\bibinfo {author} {\bibfnamefont {D.~S.}\ \bibnamefont
  {Goldbaum}}\ and\ \bibinfo {author} {\bibfnamefont {E.~J.}\ \bibnamefont
  {Mueller}},\ }\href {\doibase 10.1103/physreva.77.033629} {\bibfield
  {journal} {\bibinfo  {journal} {Physical Review A}\ }\textbf {\bibinfo
  {volume} {77}} (\bibinfo {year} {2008}),\
  10.1103/physreva.77.033629}\BibitemShut {NoStop}%
\bibitem [{\citenamefont {Umucal{\i}lar}\ and\ \citenamefont
  {Oktel}(2007)}]{Umucallar2007}%
  \BibitemOpen
  \bibfield  {author} {\bibinfo {author} {\bibfnamefont {R.~O.}\ \bibnamefont
  {Umucal{\i}lar}}\ and\ \bibinfo {author} {\bibfnamefont {M.~O.}\ \bibnamefont
  {Oktel}},\ }\href {\doibase 10.1103/physreva.76.055601} {\bibfield  {journal}
  {\bibinfo  {journal} {Physical Review A}\ }\textbf {\bibinfo {volume} {76}}
  (\bibinfo {year} {2007}),\ 10.1103/physreva.76.055601}\BibitemShut {NoStop}%
\bibitem [{\citenamefont {Umucal{\i}lar}\ and\ \citenamefont
  {Mueller}(2010)}]{Umucallar2010}%
  \BibitemOpen
  \bibfield  {author} {\bibinfo {author} {\bibfnamefont {R.~O.}\ \bibnamefont
  {Umucal{\i}lar}}\ and\ \bibinfo {author} {\bibfnamefont {E.~J.}\ \bibnamefont
  {Mueller}},\ }\href {\doibase 10.1103/physreva.81.053628} {\bibfield
  {journal} {\bibinfo  {journal} {Physical Review A}\ }\textbf {\bibinfo
  {volume} {81}} (\bibinfo {year} {2010}),\
  10.1103/physreva.81.053628}\BibitemShut {NoStop}%
\bibitem [{\citenamefont {Thomas}\ \emph {et~al.}(2017)\citenamefont {Thomas},
  \citenamefont {Barter}, \citenamefont {Leung}, \citenamefont {Okano},
  \citenamefont {Jo}, \citenamefont {Guzman}, \citenamefont {Kimchi},
  \citenamefont {Vishwanath},\ and\ \citenamefont
  {Stamper-Kurn}}]{stamperkurn2017}%
  \BibitemOpen
  \bibfield  {author} {\bibinfo {author} {\bibfnamefont {C.~K.}\ \bibnamefont
  {Thomas}}, \bibinfo {author} {\bibfnamefont {T.~H.}\ \bibnamefont {Barter}},
  \bibinfo {author} {\bibfnamefont {T.-H.}\ \bibnamefont {Leung}}, \bibinfo
  {author} {\bibfnamefont {M.}~\bibnamefont {Okano}}, \bibinfo {author}
  {\bibfnamefont {G.-B.}\ \bibnamefont {Jo}}, \bibinfo {author} {\bibfnamefont
  {J.}~\bibnamefont {Guzman}}, \bibinfo {author} {\bibfnamefont
  {I.}~\bibnamefont {Kimchi}}, \bibinfo {author} {\bibfnamefont
  {A.}~\bibnamefont {Vishwanath}}, \ and\ \bibinfo {author} {\bibfnamefont
  {D.~M.}\ \bibnamefont {Stamper-Kurn}},\ }\href {\doibase
  10.1103/physrevlett.119.100402} {\bibfield  {journal} {\bibinfo  {journal}
  {Physical Review Letters}\ }\textbf {\bibinfo {volume} {119}} (\bibinfo
  {year} {2017}),\ 10.1103/physrevlett.119.100402}\BibitemShut {NoStop}%
\bibitem [{\citenamefont {Becker}\ \emph {et~al.}(2010)\citenamefont {Becker},
  \citenamefont {Soltan-Panahi}, \citenamefont {Kronj\"{a}ger}, \citenamefont
  {D\"{o}rscher}, \citenamefont {Bongs},\ and\ \citenamefont
  {Sengstock}}]{Becker2010}%
  \BibitemOpen
  \bibfield  {author} {\bibinfo {author} {\bibfnamefont {C.}~\bibnamefont
  {Becker}}, \bibinfo {author} {\bibfnamefont {P.}~\bibnamefont
  {Soltan-Panahi}}, \bibinfo {author} {\bibfnamefont {J.}~\bibnamefont
  {Kronj\"{a}ger}}, \bibinfo {author} {\bibfnamefont {S.}~\bibnamefont
  {D\"{o}rscher}}, \bibinfo {author} {\bibfnamefont {K.}~\bibnamefont {Bongs}},
  \ and\ \bibinfo {author} {\bibfnamefont {K.}~\bibnamefont {Sengstock}},\
  }\href {\doibase 10.1088/1367-2630/12/6/065025} {\bibfield  {journal}
  {\bibinfo  {journal} {New Journal of Physics}\ }\textbf {\bibinfo {volume}
  {12}},\ \bibinfo {pages} {065025} (\bibinfo {year} {2010})}\BibitemShut
  {NoStop}%
\bibitem [{\citenamefont {Soltan-Panahi}\ \emph {et~al.}(2011)\citenamefont
  {Soltan-Panahi}, \citenamefont {Struck}, \citenamefont {Hauke}, \citenamefont
  {Bick}, \citenamefont {Plenkers}, \citenamefont {Meineke}, \citenamefont
  {Becker}, \citenamefont {Windpassinger}, \citenamefont {Lewenstein},\ and\
  \citenamefont {Sengstock}}]{SoltanPanahi2011}%
  \BibitemOpen
  \bibfield  {author} {\bibinfo {author} {\bibfnamefont {P.}~\bibnamefont
  {Soltan-Panahi}}, \bibinfo {author} {\bibfnamefont {J.}~\bibnamefont
  {Struck}}, \bibinfo {author} {\bibfnamefont {P.}~\bibnamefont {Hauke}},
  \bibinfo {author} {\bibfnamefont {A.}~\bibnamefont {Bick}}, \bibinfo {author}
  {\bibfnamefont {W.}~\bibnamefont {Plenkers}}, \bibinfo {author}
  {\bibfnamefont {G.}~\bibnamefont {Meineke}}, \bibinfo {author} {\bibfnamefont
  {C.}~\bibnamefont {Becker}}, \bibinfo {author} {\bibfnamefont
  {P.}~\bibnamefont {Windpassinger}}, \bibinfo {author} {\bibfnamefont
  {M.}~\bibnamefont {Lewenstein}}, \ and\ \bibinfo {author} {\bibfnamefont
  {K.}~\bibnamefont {Sengstock}},\ }\href {\doibase 10.1038/nphys1916}
  {\bibfield  {journal} {\bibinfo  {journal} {Nature Physics}\ }\textbf
  {\bibinfo {volume} {7}},\ \bibinfo {pages} {434} (\bibinfo {year}
  {2011})}\BibitemShut {NoStop}%
\bibitem [{\citenamefont {Windpassinger}\ and\ \citenamefont
  {Sengstock}(2013)}]{Windpassinger2013}%
  \BibitemOpen
  \bibfield  {author} {\bibinfo {author} {\bibfnamefont {P.}~\bibnamefont
  {Windpassinger}}\ and\ \bibinfo {author} {\bibfnamefont {K.}~\bibnamefont
  {Sengstock}},\ }\href {\doibase 10.1088/0034-4885/76/8/086401} {\bibfield
  {journal} {\bibinfo  {journal} {Reports on Progress in Physics}\ }\textbf
  {\bibinfo {volume} {76}},\ \bibinfo {pages} {086401} (\bibinfo {year}
  {2013})}\BibitemShut {NoStop}%
\bibitem [{\citenamefont {Tarruell}\ \emph {et~al.}(2012)\citenamefont
  {Tarruell}, \citenamefont {Greif}, \citenamefont {Uehlinger}, \citenamefont
  {Jotzu},\ and\ \citenamefont {Esslinger}}]{Tarruell2012}%
  \BibitemOpen
  \bibfield  {author} {\bibinfo {author} {\bibfnamefont {L.}~\bibnamefont
  {Tarruell}}, \bibinfo {author} {\bibfnamefont {D.}~\bibnamefont {Greif}},
  \bibinfo {author} {\bibfnamefont {T.}~\bibnamefont {Uehlinger}}, \bibinfo
  {author} {\bibfnamefont {G.}~\bibnamefont {Jotzu}}, \ and\ \bibinfo {author}
  {\bibfnamefont {T.}~\bibnamefont {Esslinger}},\ }\href {\doibase
  10.1038/nature10871} {\bibfield  {journal} {\bibinfo  {journal} {Nature}\
  }\textbf {\bibinfo {volume} {483}},\ \bibinfo {pages} {302} (\bibinfo {year}
  {2012})}\BibitemShut {NoStop}%
\bibitem [{\citenamefont {Fisher}\ \emph {et~al.}(1989)\citenamefont {Fisher},
  \citenamefont {Weichman}, \citenamefont {Grinstein},\ and\ \citenamefont
  {Fisher}}]{fisher}%
  \BibitemOpen
  \bibfield  {author} {\bibinfo {author} {\bibfnamefont {M.~P.~A.}\
  \bibnamefont {Fisher}}, \bibinfo {author} {\bibfnamefont {P.~B.}\
  \bibnamefont {Weichman}}, \bibinfo {author} {\bibfnamefont {G.}~\bibnamefont
  {Grinstein}}, \ and\ \bibinfo {author} {\bibfnamefont {D.~S.}\ \bibnamefont
  {Fisher}},\ }\href {\doibase 10.1103/physrevb.40.546} {\bibfield  {journal}
  {\bibinfo  {journal} {Physical Review B}\ }\textbf {\bibinfo {volume} {40}},\
  \bibinfo {pages} {546} (\bibinfo {year} {1989})}\BibitemShut {NoStop}%
\bibitem [{\citenamefont {Meissner}(1960)}]{meissner}%
  \BibitemOpen
  \bibfield  {author} {\bibinfo {author} {\bibfnamefont {H.}~\bibnamefont
  {Meissner}},\ }\href {\doibase 10.1103/physrev.117.672} {\bibfield  {journal}
  {\bibinfo  {journal} {Physical Review}\ }\textbf {\bibinfo {volume} {117}},\
  \bibinfo {pages} {672} (\bibinfo {year} {1960})}\BibitemShut {NoStop}%
\bibitem [{\citenamefont {Lewenstein}\ \emph {et~al.}(2012)\citenamefont
  {Lewenstein}, \citenamefont {Sanpera},\ and\ \citenamefont
  {Ahufinger}}]{tof}%
  \BibitemOpen
  \bibfield  {author} {\bibinfo {author} {\bibfnamefont {M.}~\bibnamefont
  {Lewenstein}}, \bibinfo {author} {\bibfnamefont {A.}~\bibnamefont {Sanpera}},
  \ and\ \bibinfo {author} {\bibfnamefont {V.}~\bibnamefont {Ahufinger}},\
  }\href {\doibase 10.1093/acprof:oso/9780199573127.001.0001} {\emph {\bibinfo
  {title} {Ultracold Atoms in Optical Lattices}}}\ (\bibinfo  {publisher}
  {Oxford University Press},\ \bibinfo {year} {2012})\BibitemShut {NoStop}%
\bibitem [{\citenamefont {Kuhr}(2016)}]{insitu}%
  \BibitemOpen
  \bibfield  {author} {\bibinfo {author} {\bibfnamefont {S.}~\bibnamefont
  {Kuhr}},\ }\href {\doibase 10.1093/nsr/nww023} {\bibfield  {journal}
  {\bibinfo  {journal} {National Science Review}\ }\textbf {\bibinfo {volume}
  {3}},\ \bibinfo {pages} {170} (\bibinfo {year} {2016})}\BibitemShut {NoStop}%
\bibitem [{\citenamefont {Corcovilos}\ \emph {et~al.}(2010)\citenamefont
  {Corcovilos}, \citenamefont {Baur}, \citenamefont {Hitchcock}, \citenamefont
  {Mueller},\ and\ \citenamefont {Hulet}}]{lightscattering}%
  \BibitemOpen
  \bibfield  {author} {\bibinfo {author} {\bibfnamefont {T.~A.}\ \bibnamefont
  {Corcovilos}}, \bibinfo {author} {\bibfnamefont {S.~K.}\ \bibnamefont
  {Baur}}, \bibinfo {author} {\bibfnamefont {J.~M.}\ \bibnamefont {Hitchcock}},
  \bibinfo {author} {\bibfnamefont {E.~J.}\ \bibnamefont {Mueller}}, \ and\
  \bibinfo {author} {\bibfnamefont {R.~G.}\ \bibnamefont {Hulet}},\ }\href
  {\doibase 10.1103/physreva.81.013415} {\bibfield  {journal} {\bibinfo
  {journal} {Physical Review A}\ }\textbf {\bibinfo {volume} {81}} (\bibinfo
  {year} {2010}),\ 10.1103/physreva.81.013415}\BibitemShut {NoStop}%
\bibitem [{\citenamefont {Chen}\ \emph {et~al.}(2016)\citenamefont {Chen},
  \citenamefont {Li},\ and\ \citenamefont {Su}}]{fractionalmott1}%
  \BibitemOpen
  \bibfield  {author} {\bibinfo {author} {\bibfnamefont {Q.-H.}\ \bibnamefont
  {Chen}}, \bibinfo {author} {\bibfnamefont {P.}~\bibnamefont {Li}}, \ and\
  \bibinfo {author} {\bibfnamefont {H.}~\bibnamefont {Su}},\ }\href {\doibase
  10.1088/0953-8984/28/25/256001} {\bibfield  {journal} {\bibinfo  {journal}
  {Journal of Physics: Condensed Matter}\ }\textbf {\bibinfo {volume} {28}},\
  \bibinfo {pages} {256001} (\bibinfo {year} {2016})}\BibitemShut {NoStop}%
\bibitem [{\citenamefont {Buonsante}\ \emph {et~al.}(2005)\citenamefont
  {Buonsante}, \citenamefont {Penna},\ and\ \citenamefont
  {Vezzani}}]{fractionalmott2}%
  \BibitemOpen
  \bibfield  {author} {\bibinfo {author} {\bibfnamefont {P.}~\bibnamefont
  {Buonsante}}, \bibinfo {author} {\bibfnamefont {V.}~\bibnamefont {Penna}}, \
  and\ \bibinfo {author} {\bibfnamefont {A.}~\bibnamefont {Vezzani}},\ }\href
  {\doibase 10.1103/physreva.72.031602} {\bibfield  {journal} {\bibinfo
  {journal} {Physical Review A}\ }\textbf {\bibinfo {volume} {72}} (\bibinfo
  {year} {2005}),\ 10.1103/physreva.72.031602}\BibitemShut {NoStop}%
\end{thebibliography}%

\end{document}